\documentclass[10pt]{article}
\usepackage{amsmath}
\usepackage{amssymb}
\usepackage{color}
\usepackage{graphicx}

\usepackage{cite}

\usepackage{color} 
\usepackage{epstopdf}
\AppendGraphicsExtensions{.tif}

\usepackage[labelfont=bf,labelsep=period,justification=raggedright]{caption}

\makeatletter
\renewcommand{\@biblabel}[1]{\quad#1.}
\makeatother

\date{}

\begin{document}

\begin{flushleft}
{\Large
\textbf{The causal inference of cortical  neural networks  during  music improvisations}
}
\\
Xiaogeng Wan$^{1}$, 
Bj{\"o}rn Cr{\"u}ts$^{2}$, 
Henrik Jeldtoft Jensen$^{1,\ast}$
\\
\bf{1}\textit{Department of Mathematics and Complexity \& Networks Group, Imperial College London, SW7 2AZ}, London, United Kingdom
\\
\bf{2} \textit{Brainmarker BV, Molenweg 15a, 6271ND Gulpen, The Netherlands}
\\
$\ast$ E-mail: x.wan11@imperial.ac.uk, bjorncruts@gmail.com, h.jensen@imperial.ac.uk
\end{flushleft}

\section*{Abstract}
We present an EEG study of two music improvisation experiments. Professional musicians with  high  level of  improvisation skills were asked to perform music  either according to  notes (composed music) or in improvisation. Each piece of music was performed in two different modes: strict mode  and \textquotedblleft let-go\textquotedblright  mode.  Synchronized EEG data was measured from both musicians and  listeners. We used one of the most reliable causality measures: conditional Mutual Information from Mixed Embedding (MIME), to analyze directed  correlations between different EEG  channels, which was combined with network theory to construct both intra-brain and cross-brain  networks. Differences  were identified in intra-brain neural networks   between composed music and improvisation and between strict mode and \textquotedblleft let-go\textquotedblright  mode. Particular brain regions such as frontal, parietal and temporal regions were found to play a key role in differentiating the  brain activities between different playing conditions.  By comparing the level of degree centralities in intra-brain neural networks,  we found a difference between the response of musicians and the  listeners when  comparing the different playing conditions.\\
\textbf{Key words:}\ Music improvisation, EEG, MIME, Network.

\section*{Introduction}
Improvisation, an instantaneous creative behavior,  is often encountered in different forms of art such as music and dance. In this paper, we study the brain mechanism of music improvisation. We refer to the performance according to notes as  composed music, while the instantaneous creative  performance as  improvisation. In the performances, each piece of music either composed or improvised was played in either a  mechanical manner (i.e. strict mode) or in  a more  emotionally rich manner (\textquotedblleft let-go\textquotedblright    mode).\\

Music improvisation is  believed to involve neural-substrates in large brain regions \cite{M1} \cite{M2} \cite{M3} \cite{M4} \cite{M5} \cite{M6} \cite{M7} \cite{M8} \cite{MR6} \cite{MR7} .  If these large brain regions are identified, one could use neuroscientific approaches to improve the quality of music performance. To analyze the brain mechanisms of music improvisation, we use causality measures to analyze the EEG data measured from the experiments and construct generalized neural networks for each experimental condition.  We have investigated various kinds of  causality measures, where we found the conditional Mutual Information from Mixed Embedding (MIME, a time domain direct causality measure developed by  I. Vlachos and D. Kugiumtzis \cite{MIME}) to be  the optimal causality measure for our EEG analysis. In this paper, we present our results on intra-brain and cross-brain neural networks, and compare the networks observed for different paling conditions.\\

Music improvisation has long been studied  by neuroscientists and mathematicians using various approaches.   Some recent research has identified a number of frontal brain regions, including the pre-supplementary motor area (pre-SMA)  \cite{MR1} \cite{M2}, \cite{MR2} and the dorsal premotor cortex (PMD)  \cite{MR3} \cite{MR4}, to play central roles in more cognitive aspects of movement sequencing and creative generation of music.  In recent decades, scientists have used brain scanning techniques such as  fMRI, PET and EEG to study the brain.  O. D. Manzano et. al.  used fMRI   to   study the melodic and rhythmic improvisation  in a $2\times 2$ factorial experiment  \cite{M2} \cite{M3}, where the dorsal premotor cortex (PMD, in frontal cortex, is assumed to be consistently involved in cognitive aspects of planning and selection of spatial motor sequences) was found to be the  main region for melodic improvisation,  while  the pre-supplementary motor area (pre-SMA, showing increased activation during perception, learning and reproduction of temporal sequences) was identified to be related to  rhythmic improvisation  \cite{M2} \cite{M3}.   A. L. Berkowitz et. al. \cite{M6} also used fMRI but to study the expertise-related neural differences  between musicians and non-musicians during improvisation, their results show that musicians have right temporoparietal junction (rTPJ) deactivation during music improvisation,  while non-musicians showed no activity change in this region \cite{M6}.  Moreover,  C. Babiloni et al.  \cite{M8} studied the frequency filtered EEG  measured from professional saxophonists during music performances, they found the EEG power density values  decreased in the alpha band ($8-12$Hz) in the posterior  cortex during resting state, while the  power values  enhanced within narrow high-frequency bands during   music performances \cite{M8}.  Other  studies are  e.g. the EEG phase synchrony analysis   \cite{M02},  fMRI study of jazz improvisation \cite{M1},  PET studies of melody and sentence generation  \cite{M4} and the fMRI study of pseudo-random motor and cognitive tasks  \cite{M7}. \\

Amongst  the many mathematical tools used,  we consider the the most relevant tools such as  measures of  correlations,  standardized Low Resolution Brain Electromagnetic Tomography (sLORETA, \cite{sLORETA})  and analysis of variance (ANOVA, \cite{ANOVA1}, \cite{ANOVA2}). sLORETA is a  method used to localize, identify and visualize  EEG point sources in the brain \cite{sLORETA}.  ANOVA is used to analyze the statistical differences between different experimental conditions  \cite{ANOVA1}  \cite{ANOVA2}. It has been applied to  e.g.  the music improvisation study on trained pianists \cite{M5}, which  revealed the dorsal premotor cortex, the rostral cingulate zone of the anterior cingulate cortex and the inferior  frontal gyrus are important to both rhythmic   and melodic  motor sequence creation \cite{M5}.  D. Dolan et al. undertook a sLORETA analysis of  the EEG for music improvisation \cite{DOLAN}. They use part of the same data as we have used for the analysis described in this paper, namely  EEG measurements in the trio ensemble. Their  sLORETA analysis suggested  similar results to those we obtained from the MIME analysis (see below).  In both cases the frontal cortex which was found to be key to the music improvisation \cite{DOLAN}. Other studies are e.g. the correlation analysis on either  EEG \cite{MR5} \cite{MR8} or music performance \cite{MR6} \cite{MR7}, but non of these analyses addressed the {\em direction} of causal influence between  EEG channels. In this paper, we will complement the EEG analysis in \cite{DOLAN} by using MIME and network theory.\\

For our EEG analysis, we have investigated three popular measures, namely the nonlinear indirect measures: transfer entropy (TE \cite{TE}) and  MIME  \cite{MIME},  and a linear direct measure: partial directed coherence (PDC  \cite{PDC2}  \cite{PDC3}).  TE   was found to produce unsatisfactory  resolution causality results,  while PDC returns false or unreliable causalities due to the failure of the autoregressive model fitting.  MIME appears to be the best measure for our EEG analysis, which can efficiently generate reliable and robust causality results with satisfactory resolution.  Hence,  we  only report  the MIME  analysis of the neural  information flows between large brain regions and  between the brains of musicians and listeners. Details of the comparison among causality measures will be discussed in another publication. 

\subsection*{Our study}
We consider two experiments regarding the effects of music performance for different types of music (composed music and improvisation) and different playing modes (strict mode and \textquotedblleft let-go\textquotedblright    mode). Our main interest is to identify neural information flows between EEG channels using  MIME. It is noteworthy that our emphasis was put on making the experimental environment as close as possible to real concert performances.  This emphasis was especially addressed  in the second experiment,  whose playing environment is more extensive.\\

Two music improvisation experiments were done at the Guild Hall School of Music and Drama in London separately on 20.06.2010 and 31.03.2012. Synchronized  EEG measurements from both the  musicians and the listeners were collected by Bj{\"o}rn Cr{\"u}ts  and his team (BrainMarker Corp.)  using CE-certified EEG device (Brainmarker, the Netherlands) during the music performances.\\
The first experiment consisted of the pianist David Dolan playing:
\begin{description}
\item [Test 1:] Schubert-Impromptu in G flat major Op. 90 No. 3, neutral mode, uninvolved
\item [Test 2:] Schubert-Impromptu in G flat major Op. 90 No. 3, fully  involved
\item [Test 3:] Improvisation, polyphonic, intellectual exercise
\item [Test 4:] Improvisation, polyphonic, emotional letting go.
\end{description}
The audience consisted of one listener. Both participants were connected to synchronized EEG amplifiers (250Hz sampling frequency) with 8 electrodes (P4,  T8, C4, F4,  F3, C3, T7, P3).  The electrodes are labeled by the initial of the corresponding cortices, P: parietal cortex (perception, multi-sensory integration), T: temporal cortex (processing of language and sounds), C: central cortex (sensory and motor function), F: frontal cortex (attention and executive control). The odd numbers stand for locations on the left brain, while even numbers represent locations on the right brain. The electrodes are all localized according to the international 10-20 system (Jasper, 1958). A reference electrode (Cz) at the central location on the top of the head was used, so that each EEG signal was mono-polar referenced to this central site and activity levels of the eight sites could be compared relative to each other.\\

In the second experiment, the music was performed by the Trio Anima (three highly acclaimed musicians: Drew Balch (violist), Matthew Featherstone (flutist) and Anneke Hodnett (harpist)) in the following order:
\begin{description}
\item [A. ]\  Ibert  [duration: 3\textquoteright 30\textquotedblright ]:\  1.   strict $\& $   2.  \textquotedblleft let-go\textquotedblright               
\item [B. ]\  Telemann  [duration: 2\textquoteright ]:\   1.  \textquotedblleft let-go\textquotedblright   \  $\& $  2.  strict             
\item [C. ]\  Improvisation:\  1. \textquotedblleft let-go\textquotedblright   \   $\& $  2.  strict
\item [D. ]\  Ravel   [duration: 2\textquoteright  50\textquotedblright ]:\  1. strict $\& $  2. \textquotedblleft let-go\textquotedblright  
\item [E. ]\   Improvisation:\    1. strict $\& $  2.  \textquotedblleft let-go\textquotedblright  
\end{description}
The audience con sited of about 20 listeners. Synchronized  EEG data was measured from both the musicians and from two of the the listeners, but one listener's data was excluded from the cross-brain analysis, because his data was un-synchronized with the other measured EEGs due to a  technical issue. The EEG data (100Hz sampling frequency) was measured from 10 electrodes: P4,  T8, C4, F4,  F3, C3, T7, P3,  O1 and O2 (O: occipital cortex, visual processing center).\\

 In this experiment,  pieces A, B and D are composed music (music performances according to a written score) these pieces compare with the first two tests in the first experiment. Pieces C and E were entirely improvised (instantaneous creative performance of music) by the trio, these compare with the  last two tests in the first experiment.  Both the composed music and the improvisation were played in the strict mode and the \textquotedblleft let-go\textquotedblright    \  mode.   Similar to the mode played in the test 2 and test 4 of the first experiment, the  \textquotedblleft let-go\textquotedblright    \  mode consists in a  performance with full emotional expression, whereas the strict mode is a mechanical rendition of music similar to  the neutral mode in test 1 of the first experiment. However,  the intellectual exercise (test 3) of the first experiment consist of the musician improvising a technically correct piece of music, but without any attention to emotional content. This mode  wasn't used in the second experiment.  By analyzing these experiments, we aim to identify the neural differences between the different experimental conditions.\\

 For the EEG measurements, standard EEG cap (BraiNet, Jordan Neuroscience) was used so that the electrode locations were standardized by using anatomical reference points. This ensure that measurements within and between subjects could be compared. This approach is commonly used in patient studies to compare activity levels of different large brain areas within and between subjects. Ag/AgCl electrodes with carbon shielded wires (Temec, the Netherlands) and conductive electrode gel (Ten20, D.O. Weaver$\& $Co) were used to minimize movement artifacts. Data acquisition was carried out with a sample frequency of 250 Hz in the first experiment and 100 Hz in the second experiment. Data filtering was executed using a first order 0.16 Hz high pass filter and 59 Hz fourth order low pass filter. The amplifiers were time-synchronized using a purpose build external trigger. Before the measurement the skin was cleaned using abrasive gel (NuPrep, D.O. Weaver$\& $Co.) to ensure low skin impedance ($<10$kΩ) and high signal quality.\\

 The reason for the use this 8 channel EEG recordings, is that we focus on the activities of large cortical brain regions, such as the motor cortex, rather than task-specific areas e.g. the brain regions where  finger movements are coordinated (as in event-related potential set-ups). We aim to choose an experimental set-up of minimal discomfort for the musicians but still enough electrodes to distinguish the  activity from different large brain regions. Similar  approaches have been used in other patient studies e.g. studies of  autism, where the motor cortex activity was measured. Due to the machine  set-up (active shielding mechanism), movement artifacts were minimized. Since prefrontal poles were not measured, eye movement artifacts were excluded. Similar considerations hold for other muscle activity, which most frequently  originate in the  prefrontal cortex, this region was excluded from our measurement. Muscle activity originating in the temporal regions (T) are limited to high frequencies, these frequency components ($>$32Hz) were filtered out via Fourier transforms.\\

According to Dr. David Dolan, composed music and improvisation performed in our experiments are mainly distinguished by the overall manner of the music performances \cite{DOLAN}. Improvisation  contains more coherent and long-term structural lines, shared by all members of the ensemble. The short-term beats are freer and uneven, but the deep, longer-term pulse is extremely stable in improvisation.  In composed music performances, the gestures seemed to be shorter and more rigid (even in quick repetitive phrases).  There is less room for spontaneity and audience find themselves less surprised. This is perhaps the reason behind the results of psychological tests, which showed that the improvisation is found to be more emotionally engaging and  musically  interesting by the audiences \cite{DOLAN}. Extra notes were added spontaneously  by the freer distribution of time over gestures, which leans  more significantly on structural key moments.  Another important characteristic of improvisation,  is that the risk-taking and support are provided spontaneously  by the members of the ensembles to each other. This is probably a consequence of the higher level of active listening that took place during the improvisation. Hence, one may expect that when playing  improvisation,  musicians are prevented from entering into an \textquoteleft autopilot mode\textquoteright   \ as they need to listen attentively to the music during improvisation, because the unexpected was to come.\\

 Previous research has identified a number of the frontal regions, including the pre-supplementary motor area (pre-SMA) \cite{MR1} \cite{M2} \cite{MR2} and the dorsal premotor cortex (PMD) \cite{MR3} \cite{MR4} to play central roles in more cognitive aspects  of the movement sequencing and creative generation of music. We hypothesis   that  the musicians may trigger more wide-distributed neural networks when improvising than performing composed music, and that the frontal regions (attention and executive control) play an important role in the improvisation process.\\

Given the above differences between different music types and playing modes, we aim to find their neural substrates based on the neural information flows and neural networks.  Previous music improvisation studies considered either unique point sources of the EEG  \cite{DOLAN} or  symmetric correlations between brain regions \cite{MR1} \cite{MR2} \cite{MR3} \cite{MR4}. Few studies shown neural networks constructed from the theoretic information flows between large brain regions. In this  paper, we present a causality  analysis of the music improvisation, where we aim to identify neural  differences between experimental conditions. We use the MIME causality measure to analyze the EEG data, and construct both intra-brain and cross-brain networks from the MIME causalities.  MIME is a bivariate causality measure,  the comparison between conditions are thus  based on this bivariate causality analysis rather than multivariate analysis. In consequence, the networks are constructed by using the indirect causalities. As been addressed earlier, the reason for using MIME among the many other measures is that,  it is the reliable and the most useful measure (for small systems)  for experimental data analysis. To verify the directionality of MIME in cross-brain analysis,  two reading experiments (Subsection \ref{see subsection:Causality verification of MIME}: Causality verification of MIME) was analyzed, where MIME was found to be able to identify the correct direction of cross-brain interaction from the reader to the listener.  We are convinced from this analysis that MIME does not generate false causalities and is reliable in the analysis of the music experiments.  Here,  the MIME causalities were used to construct the networks to identify the differences between experimental conditions.
 	
\section*{Results}
We have used causality measure MIME to analyze the EEG data from music experiments, from these results we construct both intra-brain neural networks between large-brain regions and cross-brain networks between musicians and listeners. 
\subsection*{Intra-brain neural information flows}
In our analysis, each  brain is considered as a neural network consisted of large cortical  brain regions connected by the neural information flows.  A link is drawn in the intra-brain neural networks if the causality value is positive and significant according to the significance thresholding test (i.e. surpass the significance threshold). To draw the links, the MIME causalities were first averaged over time windows to produce representative results. \\

\begin{figure}[ht]\centering
 \includegraphics[width=0.9\textwidth]{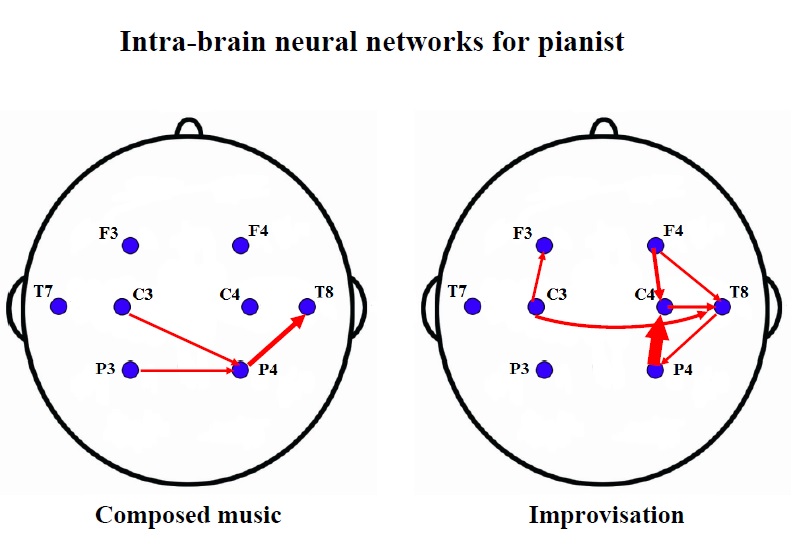}
\caption{Pianist's intra-brain neural networks  for the first experiment.  The two panels show the pianist's intra-brain neural networks separately for composed music (left) and improvisation (right). The large brain regions are labeled by the 8 electrodes:  F3, F4, C3, C4,  T7, T8, P3, P4.   The red links indicate the direction of neural information flows between large brain regions, where the thickness of the links represent the magnitudes of the causalities.}\label{fig:M_c_network and M_i_network}  
\end{figure}

\begin{figure}[ht]\centering
 \includegraphics[width=0.9\textwidth]{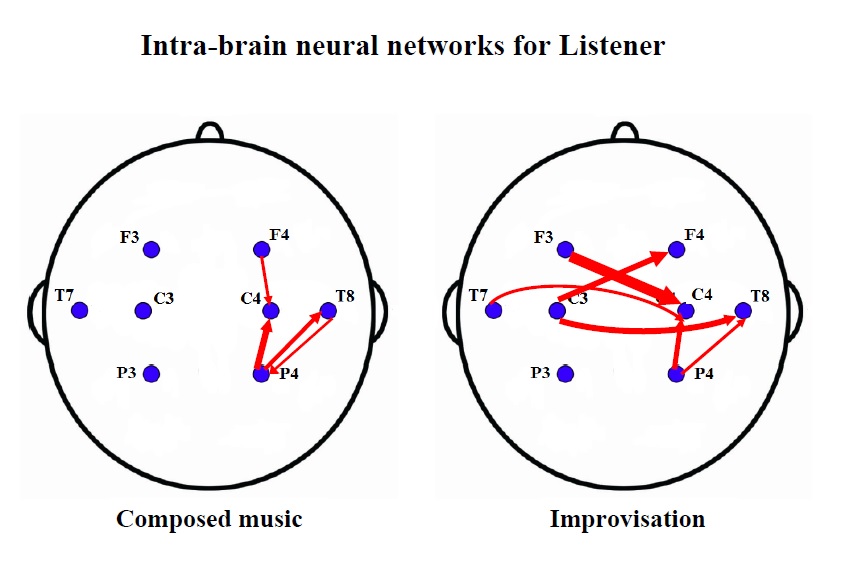}
\caption{Listener's intra-brain neural networks for the first experiment. The two panels show the listener's intra-brain neural networks separately for composed music (left) and improvisation (right). The large brain regions are labeled by the 8 electrodes:  F3, F4, C3, C4,  T7, T8, P3, P4.   The red links indicate the direction of neural information flows between large brain regions, where the thickness of the links represents the magnitudes of the causalities.}\label{fig:L_c_network and L_i_network}  
\end{figure}

In the first experiment, the intra-brain neural networks for the pianist and the listener are shown in Figures \ref{fig:M_c_network and M_i_network} (pianist) and \ref{fig:L_c_network and L_i_network} (listener), respectively.  To avoid residual information flows \cite{ResidualFlowTE}, we use a significance thresholding test to detect the significance of information flows.  The significant thresholds are taken as $T_{pianist}=0.2C_{max,pianist}$ and $T_{listener}=0.1C_{max,listener}$ for the pianist and the listener, respectively. The $C_{max,pianist}$ and the $C_{max,listener}$ are the maximum causality values (averaged over time windows) for the pianist and the listener, respectively. The specific choice of the values  $0.2$ and $0.1$ were found to produce the most clear difference between experimental conditions. From the intra-brain neural networks, the differences between composed music and improvisation were observed in the distribution of the flow of neural information. For the pianist (Figure \ref{fig:M_c_network and M_i_network}), the neural information flow is confined to the back of the brain during composed music whereas during improvisation the flow expands to the entire brain. A similarly expansion was observed for the listener (Figure \ref{fig:L_c_network and L_i_network}), although in this case the expansion is from the right brain to the entire brain when comparing composed music to improvised music.\\

 We cannot exclude that spurious causalities may exist, however the majority of the information flow is reasonable, since MIME uses a progressive scheme and a stopping criterion with significance level $A=0.95$. This level is expected to prevent false causalities while still to allow an essential causal element to be detected even if it only contributes a small amount of information. Accordingly we do expect that the causalities  predicted by the  MIME analysis is sound and that the information flows depicted in Figures \ref{fig:M_c_network and M_i_network} and \ref{fig:L_c_network and L_i_network} can be trusted. Similar interpretation of the significance applies to the second experiment.\\

\begin{figure}[ht]\centering
 \includegraphics[width=0.9\textwidth]{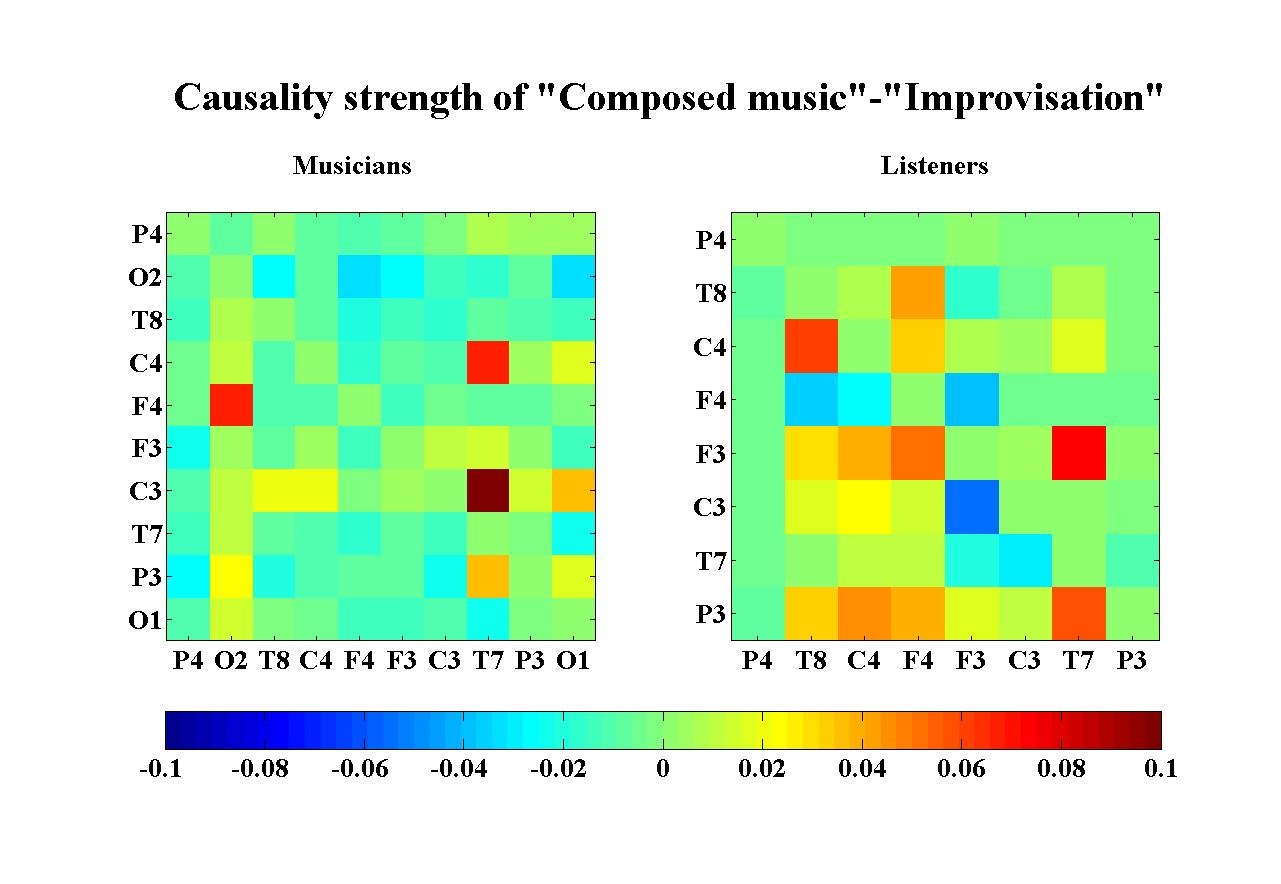}
\caption{Color-map of the contrast causality matrix between composed music and improvisation.  In this figure, the two $10\times 10$ lattice plot indicates the contrast causality matrices between composed music and improvisation separately for musicians (left) and listeners (right). The direction of information flows is   from the row channel to the column channel for each lattice. The color of the lattice indicates the strength of the causality contrasts, of difference, between composed music and improvisation, which is scaled between  $-0.1$ and $0.1$. The correspondence between the color and the causality strength is shown in the color-bar. }\label{fig:CI}  
\end{figure}

In the second experiment, an extra pair of conditions: strict mode and  \textquotedblleft  let-go\textquotedblright   mode, was added to the experiment. To compare the differences between experimental conditions, we study the contrasts,  computed as the difference, between the MIME causalities for the pairwise conditions, e.g. composed music versus  improvisation and between \textquotedblleft  let-go\textquotedblright    and strict mode.  The contrast causality values were averaged over time windows separately for the musicians and the listeners. We used a significant thresholding test to decide the significance of difference between the causalities. For each pair of conditions, e.g. the composed music vs improvisation (Figure \ref{fig:CI}), we define a radius $R=(Max_{contrast}-Min_{contrast})/2$ as the half of the difference between the global maximum and the global minimum contrasts causality values, one radius for the musicians and one for the listeners. The significance threshold was defined as half of the radius $T=R/2$, the contrast values outside the interval $(-T,T)$  were deemed significant, otherwise insignificant.  This definition of threshold is empirically reasonable,  because a lower threshold will lead to a sharp increase in the number detected of information flows, while a higher threshold will prevent reasonable direction of information flows to be registered.  For instance, the $(C3,T7)$-th lattice in left panel (musicians) of Figure \ref{fig:CI}, which has the maximum contrast values $0.1$,  indicates significant information flows from $C3\rightarrow T7$, i.e. from the left central region to the left temporal region.\\

\begin{figure}[ht]\centering
 \includegraphics[width=0.9\textwidth]{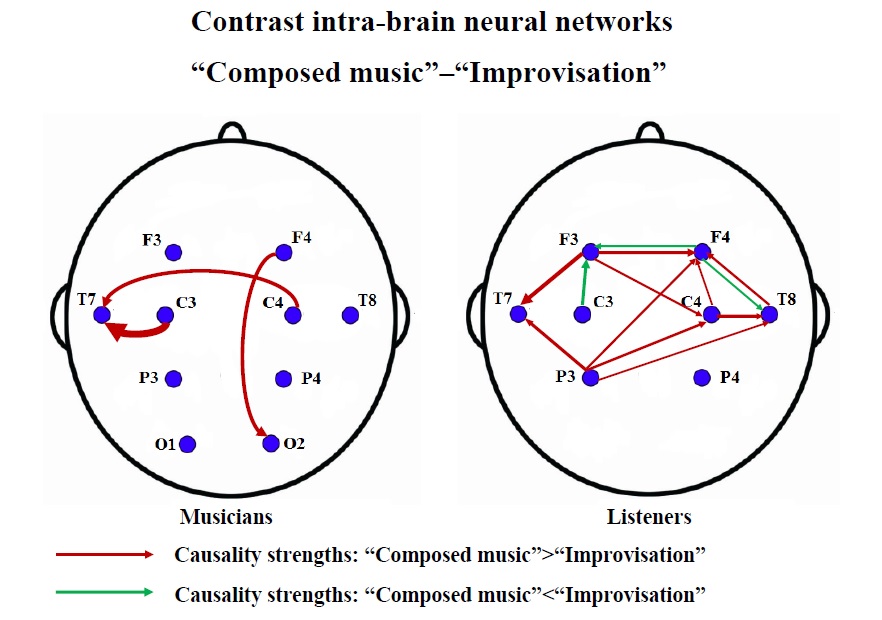}
\caption{The contrast intra-brain neural networks between composed music and improvisation for the second experiment. The contrast neural networks were drawn from Figure \ref{fig:CI}. The causality contrasts were obtained by taking the differences of the MIME causalities  between composed music and improvisation.  A link is drawn in this network if the causality contrast is significant according to a thresholding  test. The red links indicate the information flows that are significantly stronger  (causality values) in  composed music than in improvisation, while the green links indicate the information flows that are significantly stronger in improvisation than in composed music.}\label{fig:L_ci_network2 and M_ci_network2}  
\end{figure}

When the composed music is compared to the improvisation,  we find that the composed music has overall stronger intra-brain causalities than the improvisation, which is seen as more links (red) for \textquotedblleft  composed music$>$improvisation\textquotedblright  than the links (green) for \textquotedblleft  composed music$<$improvisation\textquotedblright \  in the contrast intra-brain neural networks (Figure \ref{fig:L_ci_network2 and M_ci_network2}).  This result does not contradict the observed expansion of neural information flows when composed music is changed to improvisation, this only indicates a difference in the strength of the causality values between the two conditions and corresponds to stronger values for the composed music.\\
 
For musicians (the left panel in Figure  \ref{fig:L_ci_network2 and M_ci_network2}), the significant stronger information flows of composed music are from  both the left and right central regions to the left temporal region ([C3,C4]$\rightarrow $T7) and from the right frontal region to the right occipital region (F4$\rightarrow $O2). For listeners (the right panel in Figure  \ref{fig:L_ci_network2 and M_ci_network2}),   information flows that are significant in composed music are from the left frontal and left parietal regions break into two branches, one is  to the left temporal region ([F3,P3]$\rightarrow $T7),  the other is to the right frontal  region via the right central ([F3,P3]$\rightarrow $F4, or  [F3, P3]$\rightarrow $C4$\rightarrow $F4) and right temporal regions (P3$\rightarrow $T8$\rightarrow $F4, or  [F3, P3]$\rightarrow $C4$\rightarrow $T8$\rightarrow $F4).  The left frontal (F3) and left parietal (P3) regions acted as the main sources of information flow, while the left temporal (T7) and right front (F4) regions are the main sinks, and the right central (C4) and right temporal (T8) regions serve as transit hubs. The listeners also have significant information flows during improvisation (green links in the right panel of Figure  \ref{fig:L_ci_network2 and M_ci_network2}):  from the right frontal region to the left frontal (F4$\rightarrow $F3) and right temporal regions (F4$\rightarrow $T8) and from the left central to the left frontal region (C3$\rightarrow $F3). The information flows that are significant during improvisation (red links) have directions  opposite those found during composed music (green links).   The more red links than green links in the figure of the contrast intra-brain neural networks is not in contradiction to the expansion in the distributions of information flows, when composed music is changed to improvisation. The dominance of red links only implies that those directions have stronger causality values during composed music than during improvisation. In other words,  this analysis highlights the difference in causality values between the different music types. Namely, when the flow occurs with significant different causalities for different mode of performance it will be detected in Figure \ref{fig:L_ci_network2 and M_ci_network2},  while if the information flows with more or less comparable  and significant causality values for the two conditions no contrast will be registered in this figure.\\

\begin{figure}[ht]\centering
 \includegraphics[width=0.9\textwidth]{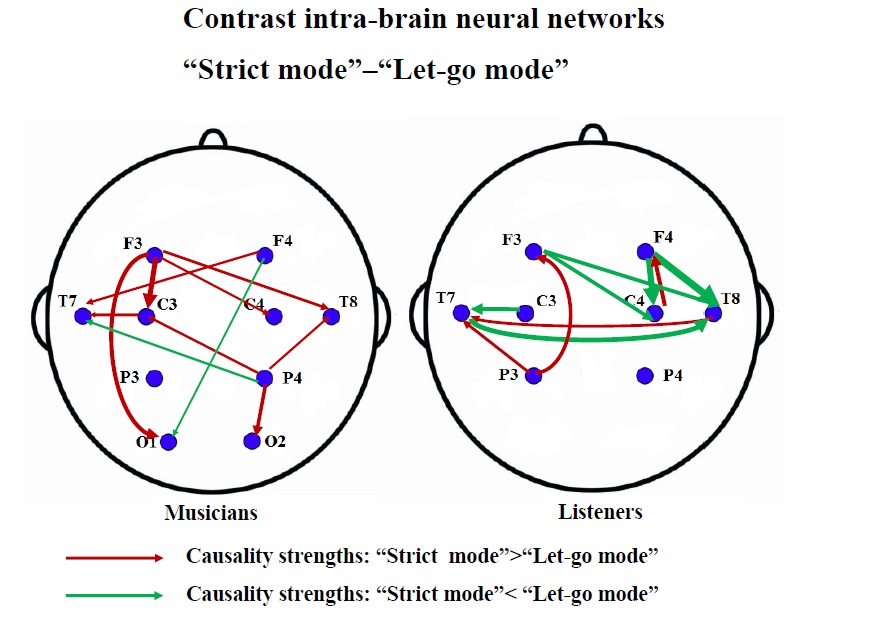}
\caption{The contrast intra-brain neural networks between strict mode and \textquotedblleft  let-go\textquotedblright  \  mode for the second experiment.  The contrast neural networks were drawn from Figure \ref{fig:CI}. The causality contrasts were obtained by taking the differences of the MIME causalities  between strict mode and \textquotedblleft  let-go\textquotedblright  \  mode.  A link is drawn in this network if the causality contrast is significant according to a thresholding  test. The red links indicate the information flows that are significantly stronger  (causality values) in  strict mode than in \textquotedblleft  let-go\textquotedblright  \  mode, while the green links indicate the information flows that are significantly stronger in \textquotedblleft  let-go\textquotedblright  \  mode than in strict mode.}\label{fig:L_sl_network2 and M_sl_network2}
\end{figure}

The network structures are more complicated for strict mode and  \textquotedblleft  let-go\textquotedblright   mode (Figure \ref{fig:L_sl_network2 and M_sl_network2}). For musicians (left panel),   information flows that are significant in strict mode are from the left frontal region to the left and right central regions (F3$\rightarrow $C3, C4) and to the left occipital (F3$\rightarrow $O1) and the right temporal (F3$\rightarrow $T8) regions, from the right frontal and left central regions to the left temporal region (F4, C3$\rightarrow $T7) and  from the right parietal region to the left central (P4$\rightarrow $C3), right occipital (P4$\rightarrow $O2) and right temporal  (P4$\rightarrow $T8) regions. Here, we see that the  left frontal region (F3) and the right parietal (P4) region are key to musicians playing in strict mode (\textquotedblleft  strict$>$let-go\textquotedblright ).   However, in the same intra-brain neural network for musicians,    information flows that are significant in \textquotedblleft  let-go\textquotedblright   mode  are from  the right frontal region to left occipital region (F4$\rightarrow $O1) and from the right parietal region to left temporal region (P4$\rightarrow $T7). For listeners, there is a clear difference in the distribution of neural information flows.  Information flows significant in strict mode (\textquotedblleft  strict mode$>$let-go mode\textquotedblright ) are from the left parietal to the left frontal (P3$\rightarrow $F3) and left temporal regions (P3$\rightarrow $T7), from the right temporal region to the left temporal region (T8$\rightarrow $T7) and from the right central region to the right frontal region (C4$\rightarrow $F4), whereas  information flows significant in \textquotedblleft  let-go\textquotedblright   mode (\textquotedblleft  strict mode$<$let-go mode\textquotedblright ) are from the left and right frontal regions to the right central (F3,F4$\rightarrow $C4)  and right temporal regions (F3,F4$\rightarrow $T8) and from the left central region via the left temporal region to the right central region (C3$\rightarrow $T7$\rightarrow $C4). In strict mode flows tend to be from the back to the front of the brain, whilst \textquotedblleft  let-go\textquotedblright   mode  trend to exhibit the inverse direction from the front to the back of the brain.\\

 Since the intra-brain  analysis studied the difference between different experimental conditions we do not have enough statistics to discuss  reliably specific instantaneous flow patterns, but have concentrated on  average trends 
 of information flows  as well as the sink and sources activities of large brain regions. These results are obtained  by averaging over time windows and over experimental conditions. 

\subsection*{Degree centrality analysis}
To identify the difference between experimental conditions in terms of the importance of large brain regions, we carried out a degree centrality analysis on the intra-brain neural networks. Since the intra-brain neural networks are directed, we use the degree centrality measure (i.e. counting the number of in-going and out-going links to a node) and separately calculate the in-degree and out-degree for each node (i.e. large brain region). The degree centralities were averaged over time windows and experimental conditions, results show that the musicians typically have opposite trends to the listeners when composed music is compared to improvisation and the strict mode is compared to the \textquotedblleft  let-go\textquotedblright   mode.\\

\begin{figure}[ht]\centering
 \includegraphics[width=0.9\textwidth]{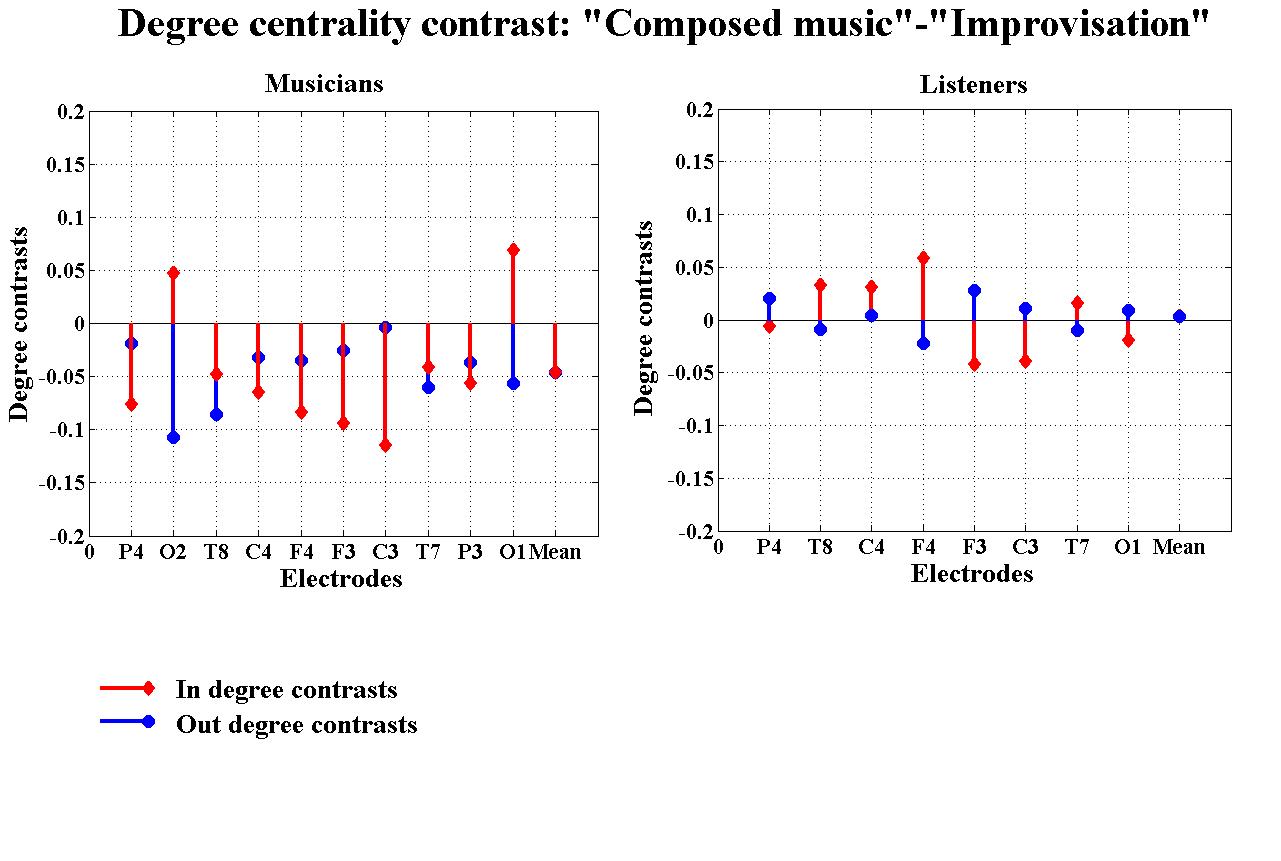}
\caption{Degree centrality contrasts, or difference, between composed music and improvisation in the second experiment. In this figure, the red stems and the blue stems indicate the in and out degree centrality contrasts between composed music and improvisation, respectively. The horizontal axis has 9 channels represent the 8 electrodes:  P4, T8, C4, F4, F3, C3, T7, P3 and the overall average over the 8 electrodes, while the vertical axis gives the magnitudes of the degree centrality contrasts between composed music and improvisation.}\label{fig:IDOD_ML_CI}
\end{figure}

\begin{figure}[ht]\centering
 \includegraphics[width=0.9\textwidth]{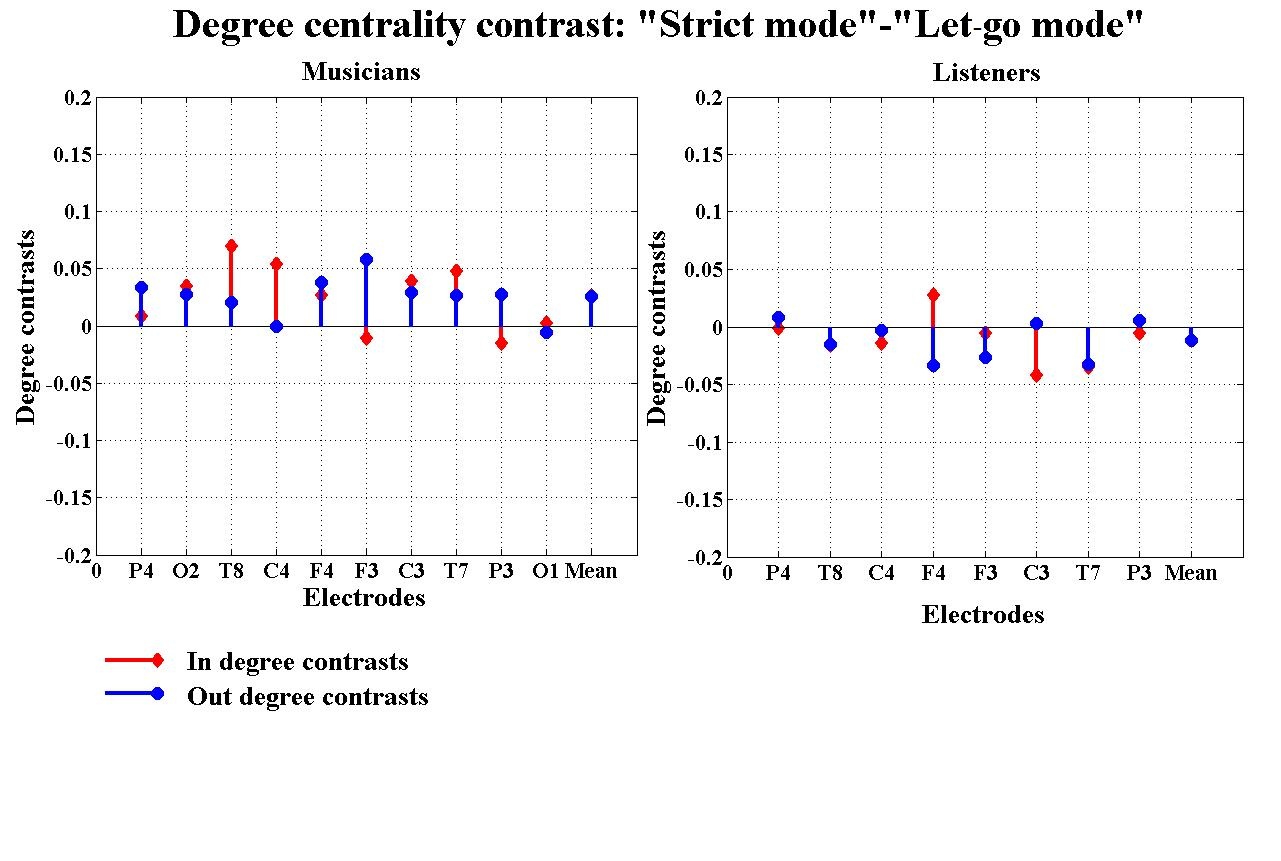}
\caption{Degree centrality contrasts between strict mode and \textquotedblleft  let-go\textquotedblright     mode in the second experiment. In this figure, the red stems and the blue stems indicate the in and out degree centrality contrasts between strict mode and \textquotedblleft  let-go\textquotedblright     mode, respectively. The horizontal axis has 9 channels represent the 8 electrodes:  P4, T8, C4, F4, F3, C3, T7, P3 and the overall average over the 8 electrodes, while the vertical axis gives the magnitudes of the degree centrality contrasts between strict mode and \textquotedblleft  let-go\textquotedblright     mode.}\label{fig:IDOD_ML_SL}
\end{figure}

The differences between the experimental conditions were compared by subtracting the degree centralities found under one condition from those found under another condition and thereby focus on the contrast between experimental conditions. In  Figure \ref{fig:IDOD_ML_CI} we show that musicians were found to have larger  in and out degrees during improvisation than during composed music, while the listeners exhibit the opposite trend. When strict mode was compared with \textquotedblleft  let-go\textquotedblright   mode (Figure \ref{fig:IDOD_ML_SL}), we find that musicians have larger in and out degrees in strict mode than in \textquotedblleft  let-go\textquotedblright   mode, while listeners again exhibit the opposite results.  In this analysis, a larger in and out degree indicates a larger amount of  information flows in and out of this brain regions and hence one would expect this to imply that  the region to be more functionally involved with the other regions in the network.

\subsection*{Cross-brain networks}
\begin{figure}[ht]\centering
 \includegraphics[width=0.9\textwidth]{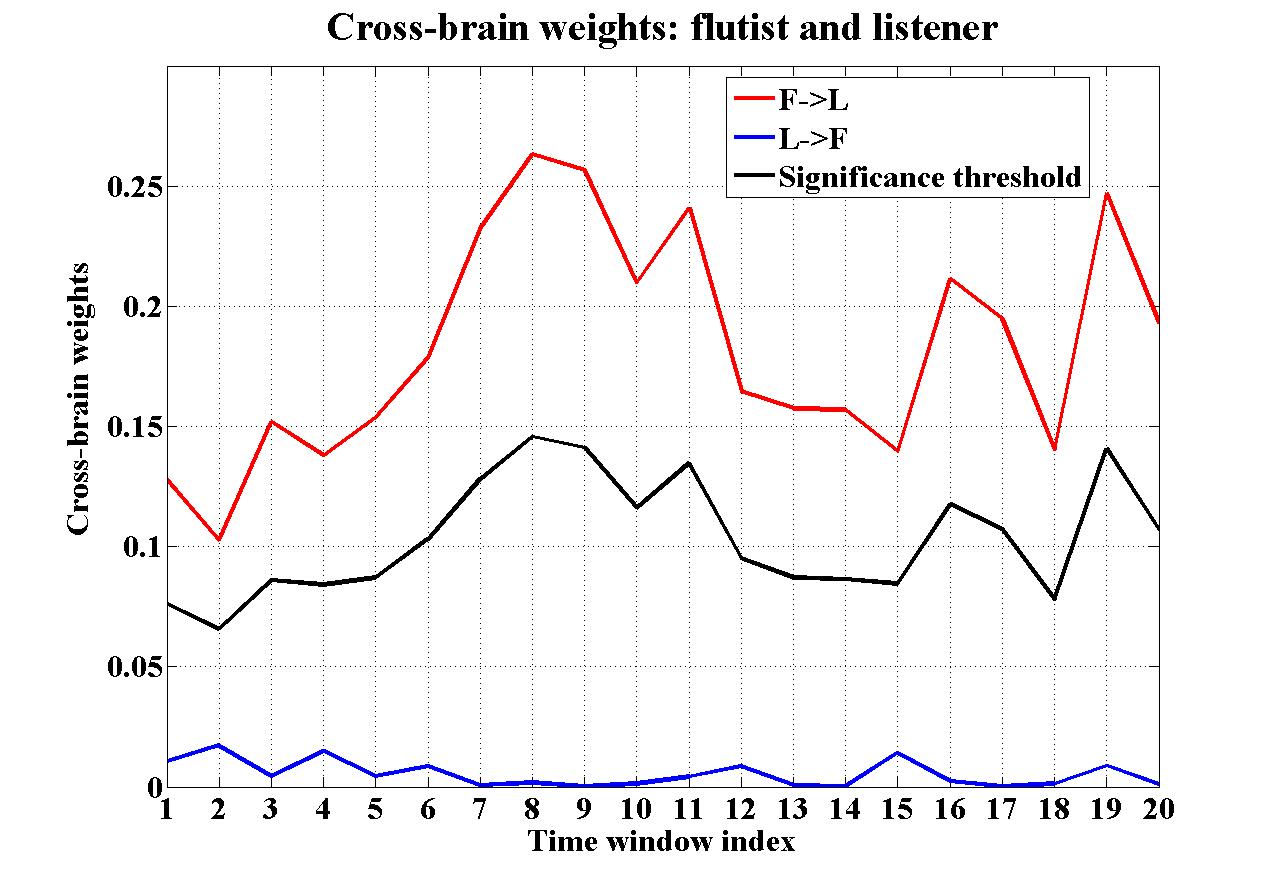}
\caption{The cross-brain weights between flutist and listener in the second experiment. This figure plots the cross-brain causalities  between flutist and listener against time windows for piece A: Ibert, strict mode.  The red curve indicates  flutist$\rightarrow $listener,  the blue curve represents listener$\rightarrow $flutist,  while the black curve is the significance threshold.}\label{fig:P1_S_FL}
\end{figure}

P. Vuust reported in \cite{Denish}, \cite{PeterVuust2} a study of jazz performances, where the jazz musicians were found to communicate with each other by modulating their individual rhythm during ensemble performances.  In our experiments we  study both the musicians and the listener, and we try to investigate the way the musicians coordinate with each other and, in terms of information flow,  interact with the listeners during the music performances.  In our study,  to analyze the pattern of coordination, we monitor the average cross-brain causalities which results in a single nonnegative real number for each direction (i.e. from one brain to another). A cross-brain link is drawn if the average causality value i.e. the cross-brain weight, is significantly higher in one direction than in the opposite direction.  For instance,  in the second experiment, the cross-brain weight is significantly higher for harpist$\rightarrow $ listener, while almost vanish for listener$\rightarrow $harpist, hence the cross-brain interaction is from the harpist to the listener during the music performances.  We do not need to define a specific threshold, since the cross-brain weight is positive with high values in one direction and almost vanishing cross-brain weight in the opposite direction between each pair of brains.  For  instance,  in Figure \ref{fig:P1_S_FL}, the cross-brain weight for flutist$\rightarrow $listener is clearly higher than  the the cross-brain weight for  listener$\rightarrow $flutist, which implies a directed link from the flutist to the listener.\\

\begin{figure}[ht]\centering
 \includegraphics[width=0.9\textwidth]{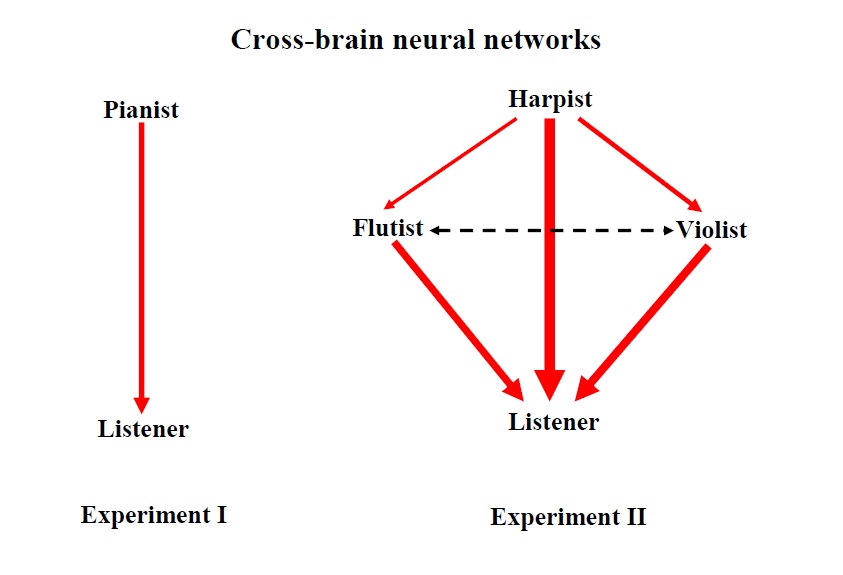}
\caption{ Cross-brain networks for the two music improvisation experiments.  The left graph is for the first experiment, while the right graph is for the second experiment. The red links represent the direction of cross-brain information flows, while the thickness of the links is proportional to the strength of the cross-brain weights (i.e. the average cross-brain causalities).}\label{fig:Crossbrain}
\end{figure}

In the first experiment (the left graph of Figure \ref{fig:Crossbrain}), the cross-brain interaction is from the  pianist to the  listener  (average weights: $A_{P\rightarrow L}=0.6554\cdot 10^{-4} > A_{L\rightarrow F}= 0.1352\cdot 10^{-4} $), while in the second experiment (the right graph of Figure \ref{fig:Crossbrain}), the cross-brain interactions are from the three musicians to the listener: [flutist, harpist, violinist]$\rightarrow $listener (average weights: $A_{F\rightarrow L}=0.1647 > A_{L\rightarrow F}=0.0304$, $A_{H\rightarrow L}=0.2002 > A_{L\rightarrow H}=0.0053$ and $A_{V\rightarrow L}=0.1901 > A_{L\rightarrow V}=0.0392$) and  from the harpist to the flutist and violinist: harpist$\rightarrow $[flutist, violinist] (average weights: $A_{H\rightarrow F}=0.0680>A_{F\rightarrow H}=0.0033$  and $A_{H\rightarrow V}=0.0945>A_{V\rightarrow H}=0.0097$).  The flutist ping-pongs with the violinist: flutist$\leftrightarrow $violinist ($A_{F\rightarrow V}=0.0509>A_{V\rightarrow F}=0.0515$, the average values are high in both directions, but the dominance of the cross-brain weights swaps between the two when the time window moves).  This network structure is robust for all performances in the second experiment.\\

 To verify the directionality of MIME in the detection of cross-brain interactions, we conducted two reading experiments. The reading experiments consist of a reader and a listener, where the reader read to the listener to establish a natural driver-responder system during the reading processes. Each reading experiment has two tests, where the reader and the listener swap their roles for the different tests.  The EEG data of the reading experiments is analyzed  by MIME to obtain the cross-brain information flows. The outcome is a causality pointing from the reader to the listener.  We mention that in one test of each experiment the cross-brain weights for the flow between the   reader and the listener were equivalent, but, importantly, both weights are insignificant,  in which case no cross-brain interaction is  detected. This is of course a limitation of the MIME analysis and shows that MIME may miss causal relations. On the other hand, the analysis of the reading experiment suggests that MIME is unlikely to produce causalities that do not exist. In other words, we believe that MIME is unlikely to produce false positives.
 
\section*{Discussion and summary}
In this paper, we constructed  intra-brain and cross-brain networks for both musicians and listeners during music performances. The differences between the composed music and improvisation and between the strict mode and \textquotedblleft  let-go\textquotedblright  \  mode can be identified in terms of the direction of neural information flows, the number of in-going and out-going connections (i.e. the in-degree and out-degree centralities) between large brain regions, as well as the sink and source activities in the frontal, parietal and temporal regions. The latter are similar to the results obtained from the  sLORETA on the same data set \cite{DOLAN}, \cite{Bjorn}.\\

In the intra-brain neural network analysis, the improvisation was found to trigger a more widely distributed network structure than the composed music did. When composed music is changed to improvisation,  the distribution of intra-brain neural information flows expands from the back of the brain to the  entire brain (for musicians), the frontal (attention and executive control) and central (motor cortex) regions become activated when musicians improvise. This may be because either performing or listening to improvisations  demands more widespread functional coordinations between large brain regions. When composed music is compared to improvisation, the intra-brain causality values are found to be greater in composed music than during improvisation,  particularly for the listeners. We find that the neural information flows start and terminate separately in left frontal and right frontal regions, the neural information flows reverse directions  when composed music is changed to improvisation and strict mode is changed to \textquotedblleft  let-go\textquotedblright   mode.  These results agree  with earlier studies \cite{MR1} \cite{MR2}:  the frontal regions (a more general area that covers the dorsal prefrontal regions), especially the right frontal region plays an important role in free improvisation of melodies and rhythms, which is the key regions that distinguish the brain activities between composed music and improvisation and between strict mode and \textquotedblleft  let-go\textquotedblright  \  mode. Moreover, the central regions tend to act as transit hubs for the neural information flows for {\em all} experimental conditions. This is in contrast to what we find for the temporal and parietal regions, which  behave differently under different experimental conditions. \\
 
 The identification of the importance of the frontal regions is similar to the findings of a previous fMRI studies of pianist improvisation \cite{M7}, where  the dorsal prefrontal cortex (part of the frontal regions) and rostral premotor regions (located within the frontal  regions) were found to be involved in the free-response selection. This study shows an activation of the cortical association areas, especially the prefrontal cortex, during divergent thinking, where the right prefrontal cortex appears to be particularly involved. The high level of involvement of the frontal and central regions and the source activity of the right frontal region during improvisation also agree with the cortical source analysis (sLORETA) on the  EEG data we have studied in this paper (see  \cite{DOLAN}). Dolan et al. found that a  clear increase in the  activation of the frontal region, acting as the EEG point sources of the brain activities, when composed music was changed to improvisation \cite{DOLAN}. Similar studies on cortical regions of music improvisation are e.g. another fMRI study on pianist improvisation \cite{M5}, which found the dorsal prefrontal and rostral cingulate regions to play a key role in melodic and rhythmic improvisation \cite{M5}.\\

In the study of intra-brain neural networks, we used the degree centrality to analyze  the level of connections between large brain regions, because the intra-brain neural networks are directed and the degree centrality measure (i.e. the number of links connected to the nodes) is very simple to use and is the most suitable centrality measure for directed networks. In this analysis, musicians were found to have opposite trends to the listeners. The musicians tend to have overall larger (in and out) degree centralities  in improvisation than in composed music, which may be because the improvisation demands more intra-brain communication for the musicians to be able to instantaneously create  melodies and rhythms. They also have larger degree centralities  in strict mode than  in \textquotedblleft  let-go\textquotedblright   mode, which may be because musicians  need more brain attention to perform in strict mode.  In contrast, the listeners have  larger degree centralities in composed music than in improvisation, which may be because the listeners found the music performed according to notes to be more familiar than the instantaneous creative performance i.e. the improvisation. The listeners also have larger degree centralities in \textquotedblleft  let-go\textquotedblright   mode than in strict mode. It is interesting to mention that a questionnaire answered by the listeners showed that music performed with free emotional expression i.e. the \textquotedblleft  let-go\textquotedblright   mode is considered more beautiful than the mechanical rendition of music i.e. the strict mode,  see  \cite{DOLAN} for details.\\

The cross-brain network structure provides a sensible view of the pattern of coordination between musicians and interactions between musicians and listeners, either during solo performance or ensembles.   In the cross-brain networks, the musicians are pointing to listeners, which seems reasonable that the musicians are communicating {\em to} the listeners during the music performances.  The harpist was frequently found to lead the flutist and the violinist, this may be an effect of  the harp   providing the chord structure which is then responded to by the flutist and the violinist during the trio's improvisation.\\ 

 We want to point out a limitation and a strength of our study. Our EEG recording has only 8 or 10 electrodes, which are quite few compared to other studies. This enables us to analyze only general brain activities. However, since we do not focus on specific task related brain  activities, this experimental set-up is sufficient for us to be able to distinguish the brain activities from different cortical brain regions and at the same time the low number of electrodes minimize the discomfort to the participants during the experiments. Nevertheless, the  results of our study imply that the neural differences in the brain of the subjects (e.g. the musicians and the listeners) under   different experimental conditions (e.g. composed music and  improvisation) can be detected by the network analysis generated from the MIME causality measures. This analysis provides a potential tool to study the intra-brain and cross-brain information flows, which is thus very promising in analyzing group behavior such as ensemble performances of music.  This method of analysis can potentially be applied to financial and more general neuroscience data sets.

\section*{Methods}\label{section:Methods}
WE use causality measures to analyze the intra-brain connectivities between large brain regions and cross-brain interactions between musicians and listeners. We have tried   three   frequently used causality measures, namely the partial directed coherence (PDC \cite{PDC2} \cite{PDC3}),  transfer entropy (TE  \cite{TE}) and conditional mutual information from mixed embedding (MIME  \cite{MIME}), in order to compare the efficiency and practicality in EEG analysis.  From the analysis, we found PDC gives large portion of cross-brain causalities  from listener to the pianist in the first experiment although listener and pianist were facing away from each other.    TE has poor directionality as  it gives similar strength for causalities between pairs of links with opposite  directions. Only MIME presents clear directionality and robust results with larger average causalities from musicians to listeners than from listeners to musicians. Therefore,  we use MIME as our core causality measure for the EEG analysis.\\

The MIME software package developed by I. Vlachos and D. Kugiumtzis, et al. \cite{MIME}  was used to calculate the causalities between EEG data  channels.  MIME is a time domain bivariate method, used to analyze nonlinear indirect information flows. It uses a progressive scheme to select  mixed embedding vectors that maximizes the conditional mutual information rate between future and past embedding vectors \cite{MIME}.  For a K-dimensional  stationary vector process $X_n=[x_{1,n},\cdots , x_{K,n}]$,  the causality from $x_j$ to $x_i$ is calculated by defining a future vector $v_F=(x_{i,n+1},x_{i,n+2},\cdots ,x_{i,n+T_i})$  containing the future of the driven variable ($x_i$), a uniform state-space embedding vector 
\[
\textbf{B}=( x_{i,n},x_{i,n-1},\cdots ,x_{i,n-L_i},x_{j,n},\cdots ,x_{j,n-L_j}) ,
\]
consists of the lagged values from both driving ($x_j$) and driven ($x_i$) variables and an empty vector $\textbf{b}_0=\emptyset $ as an initial selected non-uniform state-space embedding vector,  $T_i$ is the time horizon (prediction step) of $x_i$ and  $L_i,L_j$ are the maximum time lags for $x_i$ and $x_j$, respectively.   In each iterative cycle $s$,   the progressive scheme seeks element in $\textbf{B}\backslash \textbf{b}_{s-1}$ that satisfies the maximum criterion 
\begin{equation}\label{equation:MIMEI1}
I:\max \limits _{x_s} \{ I(v_F;x_s|\textbf{b}_{s-1})\} .
\end{equation}
which element will be add to $\textbf{b}_{s-1}$ to form a new selected vector $\textbf{b}_{s}$.  The progressive scheme stops at an $s$-th iterative circle and uses $\textbf{b}_{s-1}$ as the final embedding vector if the stopping criterion 
\begin{equation}\label{equation:MIME4}
I(x_F;b_{s-1})/I(x_F;b_s)>A,
\end{equation}
is satisfied. Here, $A \in (0,1)$ is a threshold close to 1 with empirical value $A=0.95$ (default in MIME software) for the optimum results.\\

When the progressive scheme terminates, MIME measures the causal effect from $x_{j,n}$ to $x_{i,n}$ ($i,j=1,\cdots ,K$, $i\ne j$) by evaluating the ratio between the conditional mutual information rates 
\begin{equation}\label{equation:MIME17}
MIME_{x_j\rightarrow x_i}=1-\frac{I(\textbf{v}_F;\textbf{b}_{s-1}^i)}{I(\textbf{v}_F;\textbf{b}_{s-1})}=\frac{I(\textbf{v}_F;\textbf{b}_{s-1}^j|\textbf{b}_{s-1}^i)}{I(\textbf{v}_F;\textbf{b}_{s-1})},
\end{equation}
where  $\textbf{b}_{s-1}=[\textbf{b}_{s-1}^i,\textbf{b}_{s-1}^j]$ is the final selected non-uniform state-space embedding vector when the progressive scheme terminates where  $\textbf{b}_{s-1}^i$ and  $\textbf{b}_{s-1}^j$ are the $i$th and $j$th components of $\textbf{b}_{s-1}$, respectively.\\

 Here, MIME was applied on the standardized EEG voltages.  MIME is an information based measure entirely determined by the probability distributions of the signals and therefore independent of the amplitude of the measured signal. Hence  no normalisation is necessary. To analyze cross-brain information flows, synchronized EEG data measured from each combination of two different brains was put together to form an augmented data matrix, e.g. the EEG data of the pianist and the listener. These augmented data matrices were analyzed by moving  time windows with window size $\triangle T_1=4$s ($f_{1,sample}=250$Hz)  for the first experiment and $\triangle T_2=10$s ($f_{2,sample}=100$Hz) for the second experiment. These time windowed data files were used as input to the MIME software.\\

The MIME software outputs sequences of causality matrices, which contain both intra-brain and cross-brain causalities.  These  matrices are of size $16\times 16$ ($20\times 20$) for the first (second) experiment, which consists of  two $8\times 8$ ($10\times 10$)  diagonal sub-matrices for  intra-brain causalities and  two $8\times 8$ ($10\times 10$) off-diagonal sub-matrices for  cross-brain causalities.  The diagonal sub-matrices (intra-brain) were averaged over time windows to construct intra-brain neural networks. The intra-brain causality matrices were also discretized into binary matrices, which after matrix transposition become the directed adjacency matrix for the intra-brain neural networks, the directed adjacency matrices were then used to compute the degree centralities.   The off-diagonal sub-matrices (cross-brain) were used to construct cross-brain networks by taking averages of the cross-brain causalities over electrodes and comparing the magnitudes of the causality averages with the opposite cross-brain direction. A cross-brain link is drawn from one  brain to another, if the average cross-brain causality is significantly larger from one brain to the other than measure in the opposite direction. If the average values are equivalent in both directions, the cross-brain causality cancel each other and one will not draw a link between this pair of brains.\\

 Due to the progressive scheme and the stopping criterion,  MIME behaves better than TE in terms of stability and robustness. However, similar to TE\cite{ResidualFlowTE},  MIME causalities can still have bias \cite{A7}. To overcome this limitation, we use a  significance thresholding test to decide the significance of the inferred causalities and to filter out the residual flows of  information. The thresholds are different for different analysis, details are explained  in each of the result sections. 

\subsection*{Causality verification of MIME}\label{subsection:Causality verification of MIME}
We use MIME for our EEG analysis, because it is found to be  more reliable than  TE \cite{TE} and PDC \cite{PDC2}, \cite{PDC3}. It has been tested on a number of simulations that MIME presents all correct direction inference for model data \cite{MIME} and reasonable  directional interdependencies for experimental time series such as EEG  for epilepsy patient \cite{MIME}. To our knowledge, no paper has use MIME to study music improvisation studies yet.  To verify the directionality of  MIME, we designed  two reading experiments with the aim to check the cross-brain directional inference of MIME.\\

The reading experiments include one reader and one listener, both of which are healthy normal people.  The reader is to read a short story to the listener, while the listener is to listen to the story carefully and try to imagine the scene described by the story.  When the first story  is finished. The reader and the listener swap their roles after a  short break,  to repeat the reading process on another story. The stories were new to both reader and listener. The reader and listener were prohibit to face each other during the tests, in order to avoid   visual influences. Synchronized EEG data was measured from the reader and the listener on 10 electrodes (P4, O2, T8, C4, F4, F3, C3, T7, P3, O1) during the reading processes with 100Hz sampling frequency.  The whole experiment was repeated once on another two healthy normal subjects to avoid fortuity.\\ 

The MIME analysis (time window analysis, window size: $T_{reading}=10$s) shows that the dominant cross-brain information  flow is from the reader to the listener. In the first reading experiment, the average (cross-brain) causalities are $W_{reader\rightarrow listener}=0.0523$ and $W_{listener\rightarrow reader}=0.0034$ in one test,  while $W_{reader\rightarrow listener}=0.0192$ and $W_{listener\rightarrow reader}=0.0215$ in the other test. In the second reading experiment, $W_{reader\rightarrow listener}= 0.5971$  and $W_{listener\rightarrow reader}=  0.0012$ in one case, while  $W_{reader\rightarrow listener}= 0.1008$  and  $W_{listener\rightarrow reader}= 0.1035$ in the other test.   For both experiments, a link can be drawn from the reader to the  lister, rather than the opposite direction,  because the overall causality average  is significantly greater for  reader$\rightarrow $listener than for listener$\rightarrow $reader. This is according to a significance thresholding test with instantaneous threshold $\alpha =10\%$ above the mean value between $W_{reader\rightarrow listener}$ and $W_{listener\rightarrow reader}$ at each time window. For the cases with equivalent causality strength between the reader and the listener, non of the two causal directions surpass the significance threshold, in which case it was deemed that no significant causal influence occurred between the reader and the listener. Nevertheless, the overall average for the four tests (two experiments) gives dominant cross-brain causalities from the reader to the listener.\\


We have varied the parameters of MIME, e.g.  the time horizon (prediction step) $T=1,2,3$ and the maximum embedding dimension (time lags) $L_{max}=3,4,5$ under restriction that $T<L_{max}$ \cite{MIME}. In spite of these variations  the directional results were unchanged. This implies that the directionality of MIME doesn't depend strongly on the parameter choice. Our conclusion from the reading experiments is that MIME may fail to pickup up causal links (e.g. no significant causal influence between the reader and the listener), but it never predicts an unreliable causalities. This means, once MIME picks up a  causal direction, one have good reason to believe in the directional results.

\subsection*{Selection of causality measures}\label{subsection:Selection of causality measures}
There are a number of reasons for us to use MIME in our EEG analysis. Firstly, we have compared our EEG analysis using three popular causality measures: MIME \cite{MIME}, PDC \cite{PDC2} \cite{PDC3} and TE \cite{TE}, in which MIME produces the most reliable results among  the three measures. PDC is a linear method which relies strictly on linear autoregressive models. For real EEG analysis, PDC presents larger amount of presumably false causalities from the  listeners to the musicians than from the musicians to the listeners, even though the listeners were not allowed to see the musicians and vice versa.   TE is a nonlinear method, which  is supposed to work  better than PDC in nonlinear time series analysis.  However, due to computational restrictions on embedding dimensions,  TE cannot use large enough  embedding dimensions and is for this reason unable to produce satisfactory directional results. The TE's analysis generates similar causalities between every pair of brains, so no cross-brain network structure is detected.  A small  increment  in the embedding dimension computationally very costly.\\

Both the linearity and computation short comings of other causality measures were overcome by MIME, which can produce reliable causality results efficiently \cite{MIME}.  As has been tested on various data, MIME present all correct  directional results for model data and reasonable  causalities for experimental time series \cite{MIME}.  Furthermore, we also did tests on a natural driving-driven respond system, namely the reading experiments to test the reliability of MIME in cross-brain analysis (Section: Causality verification of MIME),  from these results we concluded that MIME does not present false or unreasonable causalities when analyzing the directed interactions between experimental time series. 

\section*{Acknowledgments}
This paper acknowledges  Prof.  D. Kugiumitzis (Aristotle University of Thessaloniki, Greece) for the contribution of MIME and its software package,  international concert pianist Dr.  D. Dolan (Guildhall School of Music and Drama) and the trio Anima for music improvisation performances. We also express our gratitude to K. Mamani, J. Clough, and I. Norman for helpful comments on the manuscript. 


\section*{Figure Legends}


\begin{figure}[ht]\centering
 \includegraphics[width=0.9\textwidth]{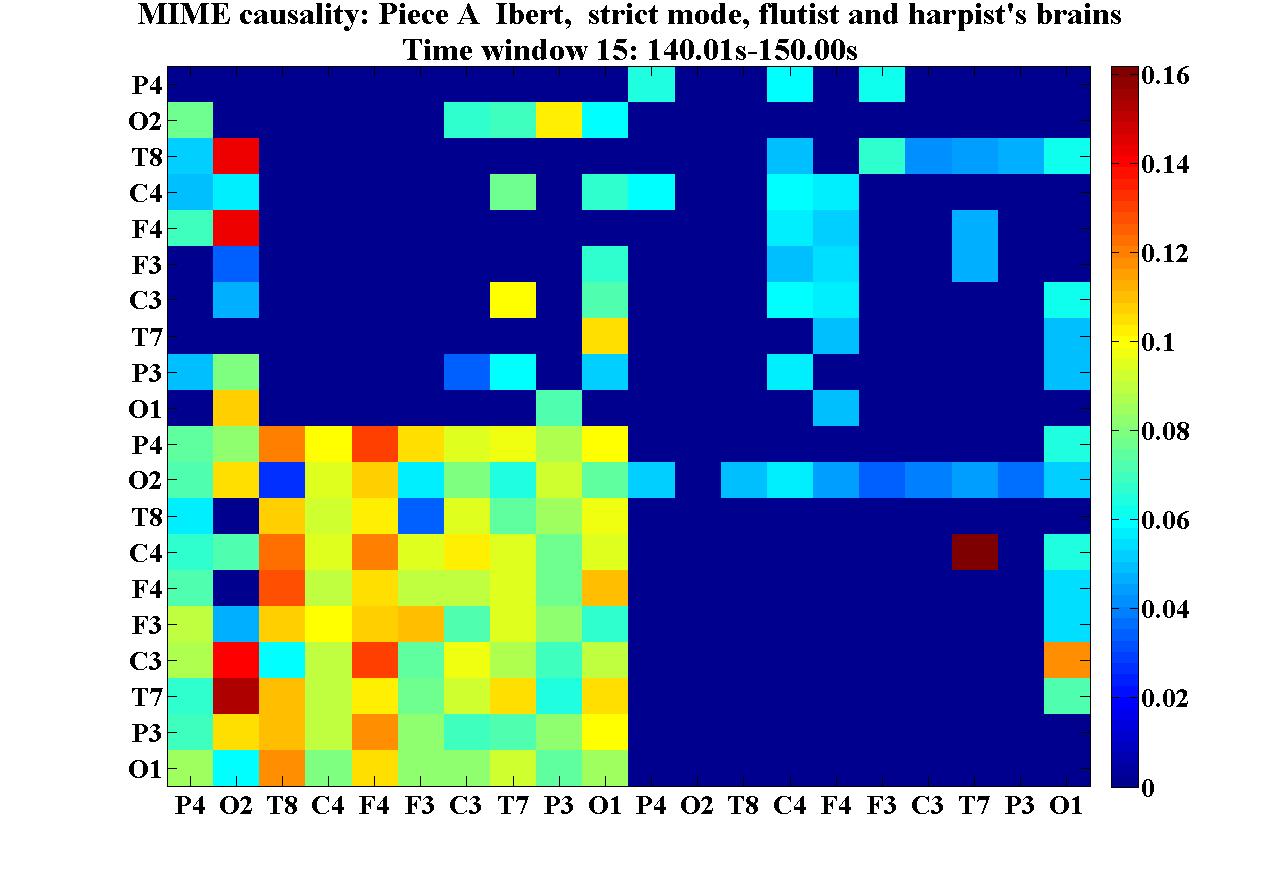}
\caption{Color-map of cross-brain causality matrix between flutist and harpist in the second experiment. This graph shows a color-map (scaled between 0 and 1) of the cross-brain causality matrix for the flutist and harpist at the time window 15 (140.01s-150.00s) during the performance of piece A: Ibert (strict mode).  The two $10\times 10$ diagonal blocks indicate the intra-brain causalities for the flutist (upper-left) and the harpist (lower-right), respectively, while the two $10\times 10$ off-diagonal blocks indicate the cross-brain causalities for flutist$\rightarrow $harpist (upper-right) and for harpist$\rightarrow $flutist (lower-left). The correspondence between the color and the causality values is shown in the color-bar.}\label{fig:fh1a_15}
\end{figure}
\end{document}